# Numerical Study of Bifurcating Flow through Sudden Expansions: Effect of divergence and geometric asymmetry

Jithin M[1*], Alok Mishra[2*], Ashoke De[2†], Malay K Das[1‡]

[1]Department of Mechanical Engineering, [2]Department of Aerospace Engineering

Indian Institute of Technology Kanpur, Kanpur, UP 208016, India

**Abstract:** A numerical study of laminar flow through symmetric and slightly asymmetric sudden expansion, of expansion ratio 1:3, in channels with increasing cross section, is carried out using two different approaches: Conventional CFD and Lattice Boltzmann Method. The effect of divergence of walls of the channel, after a sudden expansion, on the symmetry of flow and recirculation is studied for various Reynolds numbers. It is seen that the angles of the walls play an active role in disrupting the symmetry of flow. For non-parallel walls, the symmetry breaking bifurcation phenomenon no longer exists and the loss of symmetry is a gradual process. The effect of asymmetry of geometry on flow is also studied by considering two types of asymmetry: First type is at the plane of expansion where the steps on either side are of unequal heights, while the second one deals the walls of the channel are at a different inclination with the direction of inflow. The present study reveals that small asymmetry impedes the sharp transition and thus 'smoothens' the pitchfork bifurcation. Increased asymmetry finally leads to the unique solution of the governing Equations, indicating complete loss of bifurcation pattern.

---

[*] Graduate student
[†] Assistant Professor, Corresponding author, E-mail: ashoke@iitk.ac.in, Ph: +915122597863, Fax : +915122597561
[‡] Associate Professor

# 1. Introduction

Fluid flow through a symmetric sudden expansion constitutes one of the classical problems in fluid mechanics [1]. While the geometry seems quite simple and is close to that of the plain Poiseuille flow, the fluid motion in a sudden expansion is surprisingly complex and suggests multiple solutions [2]. Experiments show that the steady flow through a symmetric sudden expansion undergoes the symmetry-breaking bifurcation in the laminar two-dimensional regime. Increasing Reynolds number (Re) yields further sets of bifurcations leading to increasingly intricate recirculation patterns [3]. The steady, two-dimensional flow eventually develops unsteadiness and three-dimensionality leading to turbulence. Owing to the bifurcation-rich flow features, flow through sudden expansion acts as a test bed for hydrodynamic stability theories. Besides, sudden expansion geometry is quite common in a variety of engineering applications including the entry regions of fuel cells, reactors, and heat exchangers. The geometry plays important role in applications like burners and cyclone separators. Especially, control of flow through sudden expansion geometry remains an important issue in many devices of engineering importance [4].

A number of studies have been done in relation to sudden expansion geometries. A few of the important ones are referred here. Notable work was done by Durst and his coworkers [5] who used experimental and numerical techniques to investigate the bifurcation of laminar flow through a symmetric sudden expansion. For low Re, the study reported symmetric separation region behind the expansion. When Re exceeds a critical value, the separation region turns asymmetric, such that one recirculation length exceeds the other. While the critical Re depends on the expansion ratio, two stable solutions are obtained immediately beyond the critical Re, one being the mirror image of the other. Fearn et al. [6] showed that the sudden appearance of asymmetry requires perfectly symmetric geometry, while small geometric asymmetry produces gradual emergence of the asymmetric flow.

Tutty and Pedley [7] reported another variation of the geometric asymmetry introducing small divergence in the downstream. The study showed that, depending on the angle of divergence and the Re, the recirculation pattern may evolve as that in a Jeffery-Hamel type flow. The study also underscores the limitation of linear stability theories in predicting the growth of disturbances. Hawa and Rusak [8] also studied the effects of slight geometric asymmetry for laminar flow through sudden expansion. An asymptotic stability analysis shows the sensitivity of critical Re on the geometric asymmetry.

Mizushima and his coworkers [9] studied the hydrodynamic stability of a variety of sudden-expansion flows. For instance, when a sudden contraction follows the expanded part the flow field undergoes a series of bifurcations depending on the Re and the aspect ratio. The initial symmetric flow becomes asymmetric at a critical Re before being symmetric again at a higher Re. The two pitchfork bifurcations are then followed by a Hopf bifurcation leading unsteady flow. For lower aspect ratio, however, the flow averts the pitchforks and the steady symmetric flow field directly develops unsteadiness [10]. Linear and weakly nonlinear stability analyses predicted the presence of hysteresis between two pitchfork bifurcations and the subsequent exchange of instability modes.

In spite of the availability of a wealth of literature for flow through symmetric sudden expansion, fluid flows through slightly asymmetric channels remain largely uninvestigated. Such a study is important since perfect symmetry remains elusive in most engineering applications. Furthermore, in a bifurcating flow, a small degree of asymmetry may produce a profound effect on the flow structure. Additionally, knowledge of the effect of asymmetry can even be utilized to control the flow in sensitive applications. The present study attempts to address this issue via comparison of flows through symmetric and slightly asymmetric sudden expansion geometries. Simulations for symmetric sudden expansion into diverging channels are investigated and the effect of the rate of divergence of the walls is studied. Furthermore, two different asymmetries, as shown in Fig 1, are considered in the present studies. The first study comprises the $\zeta \neq 0$ case that involves nonalignment of the centerline axis of the inlet section to that of the expanding channel. In the second study, two walls of the expanding channel diverge at different angles from the inflow direction.

Another novelty of the present study is the use of both conventional CFD approach and lattice Boltzmann method (LBM) to assess the predictive capability of both approaches. Although LBM showed significant success in fluid flow simulations, use of LBM in bifurcating flows is surprisingly limited. He et al. [11] successfully applied LBM in simulating flow through sudden expansion and discussed the issues of using LBM in such simulations. The authors showed that in LBM formulation, Re can be varied in two distinct ways and the challenge is to keep the final results independent of the way the Re is varied. The authors have simulated the flow through sudden expansion and proposed the techniques of accurate simulation in non-uniform grids. Luo [12] tested LBM approach for simulating flow through a channel with a sudden expansion and a subsequent contraction. Zarghami et al. [13] used a hybrid finite-volume LBM to predict the flow bifurcation in a suddenly expanded channel.

Available literature shows studies in three different geometries namely, symmetric sudden expansion, slightly asymmetric sudden expansion and symmetric expansion with contraction. For brevity, present simulations cover only the first two cases while exploring multiple asymmetric-expansion geometries. Moreover, the $\zeta \neq 0$ asymmetry is considered only for cases in which both walls are at $\alpha = 0$ the position. For the second type of asymmetry, it is assumed that one wall is aligned with the inflow direction while the inclination of the other wall varies. An additional case is also considered where the two types of asymmetries are superimposed. For all these cases special attention is given to study the change in symmetry of flow and the variation of recirculation lengths with geometry.

## 2. Numerical Procedure

### 2.1 Navier-Stokes modeling

The non-dimensional governing equations for viscous Newtonian incompressible flows are given by equation (1) and (2)

$$\text{Continuity equation: } \frac{\partial u_j}{\partial x_j} = 0 \tag{1}$$

$$\text{Momentum equation: } \frac{\partial u_i}{\partial t} + u_j \frac{\partial u_i}{\partial x_j} = -\frac{1}{\rho} \frac{\partial p}{\partial x_i} + \nu \nabla^2 u_i \tag{2}$$

Where $u_j$ is Cartesian component of the velocity vector, $x_j$ is Cartesian coordinate vector component and $\nu$ is Kinematic Viscosity.

### 2.2 Lattice Boltzmann method modeling

As mentioned before, the present study also uses LBM simulator to compute flow through expanding channels. In recent years, Lattice Boltzmann method (LBM) has evolved as a promising alternative to the traditional computational fluid dynamics (CFD) techniques [14]. LBM refers to a class of computational techniques that solves the Boltzmann Equation for an ensemble-averaged distribution of interacting particles over a symmetric, discrete lattice [15]. The particle distribution function is then correlated to the macroscopic variables such as velocity and pressure. Unlike the usual Navier-Stokes approach, the LBM linearizes the convection operator and uses an Equation of state for the pressure calculation. Again, the LBM uses a minimal set of velocities in phase space as opposed to a complete phase space of the traditional kinetic theory. The LBM-based approach, therefore, selectively fuses various features of the molecular- and continuum-scale transport models for practical problem-solving [16]. In fluid mechanical applications, LBM has been successfully applied to a variety of classical problems including driven-cavity flow, backward-facing step and flow past cylinders [17]. LBM studies are also extended for a wide range of low and high Reynolds number (Re) flows [18]. While low Re indicates diffusive flow, high Re leads to increased nonlinearity, i.e., the convection-dominated flow. Present research involves a class of flows where geometric inhomogeneity augments the nonlinear effects [19]. Particular focus, in the present investigation, is the LBM simulation of laminar flow through a symmetric or slightly asymmetric sudden expansion [20].

Present work uses a D2Q9 single relaxation BGK Lattice Boltzmann formulation [21] for which the streaming and collision steps can be shown as:

$$\text{Streaming: } f_\alpha \left( x_i + c_\alpha \Delta t, t + \Delta t \right) = \tilde{f}_\alpha \left( x_i, t \right) \tag{3}$$

$$\text{Collision: } \tilde{f}_\alpha \left( x_i, t \right) = f_\alpha \left( x_i, t \right) - \frac{1}{\tau} \left[ f_\alpha \left( x_i, t \right) - f_\alpha^{eq} \left( x_i, t \right) \right] \tag{4}$$

Here $\Delta t$ and $c_\alpha \Delta t$ are time and space increments and $f_\alpha$ and $c_\alpha$ are the particle velocity distribution function and velocity component in the $\alpha$ direction respectively. $f_\alpha^{eq}$ is the equilibrium distribution function and $\tau$ is the single relaxation time. A D2Q9 model is selected with the velocity components being defined as:

$$c_\alpha = ce_\alpha = \begin{cases} (0,0) & \alpha = 0 \\ c\left[\cos\left(\frac{\alpha-1}{2}\pi\right), \sin\left(\frac{\alpha-1}{2}\pi\right)\right] & \alpha = 1,2,3,4 \\ \sqrt{2}c\left[\cos\left(\frac{\alpha-5}{2}\pi + \frac{\pi}{4}\right), \sin\left(\frac{\alpha-5}{2}\pi + \frac{\pi}{4}\right)\right] & \alpha = 5,6,7,8 \end{cases} \quad (5)$$

The expression for the equilibrium function used is given by:

$$f_\alpha^{eq} = \rho w_\alpha \left[1 + \frac{3}{c^2} c_\alpha \times u + \frac{9}{2c^4}(c_\alpha \times u)^2 - \frac{3}{2c^2} u \times u\right] \quad (6)$$

Here $w_0 = 4/9$, $w_1 = w_2 = w_3 = w_4 = 1/9$ and $w_5 = w_6 = w_7 = w_8 = 1/36$. The macroscopic density and velocity are related to the distribution function by $\rho = \sum_1^9 f_\alpha$ and $\rho u = \sum_1^9 f_\alpha c_\alpha$. The relaxation time and the viscosity are related by $\upsilon = \frac{(\tau - 0.5)c^2 \Delta t}{3}$.

## 3. Computational Domain, Validation, and Grid Independence Test

The 2-D computational domain, shown in Fig 1(a) and (b), comprises a channel inlet section of height *h* that expands to a height of (3-ζ) *h* where ζ=0 represents the symmetric sudden expansion and a small value of ζ exhibits slight asymmetry. In general, the walls, in the expanded section, may not be parallel and make an angle *α* with the *x*-axis. Nonzero values of *α* show an expanding-diverging channel, which remains symmetric if the values of *α* remain same for both walls, asymmetric otherwise. The channel extends to a length of 60*h*.

### 3.1 Navier-Stokes modeling

Only Symmetric sudden expansion channel with α = 0° has been utilized for Navier stokes modeling to assess the predictive capability of LBM while compared with the measurements and conventional CFD data. A fully develop flow is specified at the inlet for velocity, based on channel height and Reynolds number while the zero gradient is applied to pressure. At outlet fixed value zero and advective are specified to velocity respectively. No-slip boundary conditions are enforced at top and bottom wall.

The flow governing equations are solved in OpenFOAM framework [22]. The standard incompressible solver pisoFoam is utilized for laminar flow modeling, based on the finite-volume (FVM) factorized method. Pressure and velocity coupling are handled using PISO (Pressure Implicit with splitting of Operators) algorithm. Second order discretization schemes are used for all the terms. An adjustable variable time integration is used to guarantee the local Courant number is less than 1 (CFL<1).

2-D dimensional computational grids used in the current simulations are generated using ANSYS® ICEM-CFD [23]. Grid comparison has been performed with three different grids for Reynolds number 40 which are shown in Table 1. It can be observed from Fig. 2(d) that by increasing the no. of grid points, there seems to be no significant difference in simulated data as compared to the results obtained by the coarse grid. Hence, the coarse grid has been selected for further analysis.

Table 1: Computational Grids for NS-Simulation

| Grid | Grid Size |
| --- | --- |
| Coarse | 141,516 |
| Medium | 322,366 |
| Fine | 668,446 |

Velocity profile of baseline sudden expansion channel has been compared with the experimental measurement reported in Fearn et al. [6] at different x/h along the length of the channel for Reynolds number 40, 80 and 187 as shown in Fig 2(a), 2(b) and 2(c), respectively. It can be observed that the present simulated results provide a good prediction of flow fields in all the cases, similar to the simulation of Battaglia et al. [24] and Hawa and Rusak [8].

**3.2 Lattice Boltzmann modeling**

As used for N-S simulation, here also the inlet velocity profile remains parabolic with an average value of $U_{avg}$. Bounce-back [25] and zero-gradient conditions are imposed at the solid wall and the outlet, respectively. In the present formulation, the Re, relaxation time $\tau$ and skin friction coefficient $C_f$ are defined as

$$\mathrm{Re} = \frac{U_{avg} h}{\upsilon}; \quad \tau = \frac{3 U_{avg} h}{\mathrm{Re}} + \frac{1}{2}; \quad C_f = \frac{2\left(\rho \upsilon \frac{\partial u}{\partial y}\right)_{wall}}{\rho_{in} U_{avg}^2} \quad (7)$$

where $\upsilon$ is the kinematic viscosity, $\rho$ is the local density and $\rho_{in}$ is the density at the inlet.

In the present research, some of the simulations are carried out for sloping wall geometry, as can be seen in Fig 1. The boundary conditions on the sloping walls are implemented as per the formulation proposed by Guo et al. [26]. In this technique, the wall boundary is assumed to be located between a solid and a fluid node while the streaming step is decomposed in equilibrium and non-equilibrium parts. This approach has been successfully employed for various problems of complex geometric shapes [27].

The computer program, developed for the present simulation, has been extensively validated with the published results [28-30]. Figure 3 (a), (b) and (c) compares velocity profiles, obtained from the present simulator, for 2-D laminar flow through symmetric sudden expansion at Re=40, 80, 187 with the published results [6, 8, 24]. Present simulator captures the transition Re with about ±1% accuracy. Maximum deviation in the velocity is found to be less than 2%.

A grid independence test is carried out and the velocity profiles are observed to be independent of the grid at h=30 lattice units and finer grids as shown in Fig 3(d). The following factors are taken care of while selecting the grid size: for a fixed flow velocity, when Re is increased, the value of $\tau$ decreases. Very low values of $\tau$ ($\approx 0.5$) can lead to instability. In addition, proper care is taken to keep the flow velocity as low as possible in order to keep the compressibility error negligible. Keeping this in mind, we adopt the grid size of h=50 lattice units and $U_{avg}$=0.05 lattice units. It is also observed that the variation in recirculation length between h=50 case and h=70 case is around 0.04%, from which it is assumed that the results are grid independent.

## 4. Results and Discussion

### 4.1 Bifurcation prediction for Symmetric sudden expansion with α = $0^0$

In this section, the symmetry breaking behavior is discussed using both the computational approach adopted herein. Since the flow through sudden-expansion channel experiences bifurcation and give multiple solutions as the Re increases. One of the most important features of this kind of flow is to predict the critical Re at which the bifurcation takes place. That gives a lot of confidence in the numerical method for the detailed analysis.

### 4.1.1 Navier-Stokes Simulations

Bifurcation diagram (Fig.4) is obtained using the method used by Fearn et al. [6]. Here a nonzero value of the *y*-direction velocity (*V*) at the geometric centerline of the channel indicates the flow bifurcation. The Re used in Fig 4 is defined on the basis of inlet maximum velocity, $U_{max}$, and inlet section half height as mentioned in the literature.

It has also been observed through velocity validation (Fig 2) that flow is symmetric for Re=40 and it is asymmetric for Re 80 and 187. It can be also concluded that symmetry breaking bifurcation has taken place between Re=40 and 80. Then, series of simulations has been carried out for baseline sudden expansion channel, and it has been observed that the flow remain symmetric up to Reynolds number 53 which also can be observed from the bifurcation diagram as depicted in Fig. 4. Similar Navier-Stokes simulations have been carried out in many other studies which are shown in table 2. Notably, all the published literature show similar behavior including our present study, and in fact, our simulation exhibit quite a good match with the results of Hawa & Rusak [8] while the obtained critical Reynolds number is around 53.

Table 2: Symmetry-breaking bifurcation through different studies

|  | Expansion Ratio | Method | Critical Re |
| --- | --- | --- | --- |
| Shapira et al. [31] | 1:3 | Bifurcation calculation | 55 |
| Battaglia et al. [24] | 1:3 | Experiments | 44 |
| Battaglia et al. [24] | 1:3 | Simulation | 57-58 |
| Battaglia et al. [24] | 1:3 | Stability analysis | 53.8 |
| Hawa and Rusak [8] | 1:3 | Simulation | 53.8 |
| Fern et al. [6] | 1:3 | Simulation | 40.5 |
| **Present N-S Study** | 1:3 | **Simulation** | **53** |
| **Present LBM Study** | 1:3 | **Simulation** | **40.7** |

**4.1.2 Lattice Boltzmann Simulation**

Compared to the N-S simulation, the present LBM simulations predict the critical Re to be 40.7 which is found to be in excellent match with the reported value of 40.5 [6]; whereas Battaglia et al. [24] reported three different sets of critical Re: (a) 57-58 from their numerical simulations, (b) 53.8 from the stability analysis and (c) 44 from experiments. It can be depicted from the above analysis that the Lattice Boltzmann simulations provide a better prediction as compared to Navier-Stokes simulations. It can also be concluded though table 2 that many studies have been done for sudden expansion channel with expansion ration 1:3 and they have found bifurcation at around Re = 53. The present LBM simulations provide the closest value of symmetry-breaking bifurcation as compared with experimental data of Fearn et al. [6], which shows the superior predictive capability of LBM in simulating this kind of flows. Hence, LBM approach has been considered for the rest of the detailed study.

**4.2 Symmetric sudden expansion**

Baseline simulations are conducted for $\alpha$ values of $0^0$, $1^0$, $2^0$ and $5^0$ for Re of 40, 80 and 120. All the cases provide steady final solutions except Re=120 case for $\alpha=5^0$ where the flow becomes unsteady The velocity and vorticity contours are shown in Figures 6, 7 and 8, and the distribution of skin friction coefficients along the top and bottom walls for $\alpha=0^0$ are shown in Fig 5(a). It is evident from the plots that flow is symmetric for Re=40 but not so for Re=80 and Re=187.

The exact positions of primary and secondary recirculation bubbles and their lengths are obtained from the Fig 5(a). The primary recirculation bubbles for Re=40 on the top and the bottom walls are almost same and the recirculation length is observed to be 4.08h on the top wall and 4.07h on the bottom wall from the plane of expansion. Whereas for Re=80, the larger primary recirculation bubbles extend to a distance of about 10h from the expansion, while the length of the smaller one is reduced to around 3.6h. Even though the $C_f$ variation along the lower wall shows a tendency to decrease after the region of the recirculation bubble, it never reaches zero again revealing the fact of non-existence of secondary recirculation bubble for Re=80. While looking at Re=187, the larger of the recirculation bubble has a length of 15.25h approximately from the plane of expansion, whereas the smaller primary recirculation bubble length is 4.46. The $C_f$ variation on the lower wall after changing sign from negative to positive again goes down to negative, between $x$=12.6h to about 26h, showing the presence of a secondary recirculation bubble for Re=187 [Fig. 5(a)]. The $C_f$ distribution also provides information regarding the distance after which the flow becomes fully developed. For instance, for Re=40, the flow develops after 20h whereas for Re=80, almost 40h distance is required. In the case of Re=187, the flow develops just before 60h which shows that that as Re increases larger distances are required for the flow to be free from the effect of the expansion. For Re greater than 187, the flow may become completely unsteady and three-dimensional, which is beyond the scope of the present work and can be considered as a part of future work.

Figure 5(b) depicts the variation of primary and secondary recirculation lengths with Re for $\alpha$=0 (symmetric expansion). It is clearly observed that the primary recirculation length at one wall increases while that on the other wall decreases after the bifurcation point. The value of critical Re is also obtained from this plot and the value turns out to be ~40 which conform to the previous results. As the Re is further increased, secondary recirculation zones start showing up along the same side of the wall on which smaller primary recirculation bubble exists. This happens at a Re of around 76. Figure 5(b) also exhibits the distance from the plane of expansion at which the secondary recirculation starts and ends for different Re.

**4.3 Effect of divergence, i.e. $\alpha$ variation**

In this section, the variations in a flow field with $\alpha$ is investigated from simulations at Re=40, 80 and 120. Skin friction distribution along the top and the bottom walls are plotted for better observation of the flow features as shown in Figures 9(a), 9(b) and 9(c). The solid line shows the $C_f$ distribution along the top wall and the symbols show the $C_f$ distribution along bottom wall.

From Re=40, it can be observed that for a diverging channel, the $C_f$ increases just after the recirculation region and then decreases continuously. It is observed that increase in α results in the decrease of $C_f$. Symmetry breaking is also observed as α is increased for fixed Re. This can be seen from the figure as the $C_f$ of top and bottom walls no longer overlap for α=$5^0$ unlike the smaller α values. Figure 9(b) shows the $C_f$ plots for Re=80 in which also similar patterns are observed. The flow, in this case, is asymmetric for all values of *α*. The recirculation regions elongated as *α* is increased. But as the flow becomes free from the influence of the sudden expansion, $C_f$ is observed to decrease with *α* . Also, a secondary recirculation bubble is formed at *α*=$2^0$. For *α*=$5^0$ case, the behavior of the $C_f$ curve is similar but the asymmetry is more profound. Re=120 case also shows a behavior similar to Re=80 (Fig. 9(c)). But in the case of Re=120 and *α*=$5^0$, the $C_f$ distribution curve is not smooth, unlike the other cases. This shows that the flow becomes unsteady for this case which needs detailed unsteady analysis.

To find how the critical Re of symmetry breaking varies with *α*, asymmetry values are plotted for different Re (Fig 9(d)). Here asymmetry is measured as the sum of the absolute value of vertical velocity at each point on the horizontal axis through the center of the channel. The curves for different *α* exhibit the similar behavior of symmetry breaking with Re. It can be easily seen that for non-parallel wall cases, there is no critical Re of symmetry breaking as the loss of symmetry is observed to be gradual. Also, as *α* increases, the loss of symmetry takes place at lower Re. Stability analysis might be required for better understanding of these behaviors, which builds the scope for future studies.

## 4.4 Effect of asymmetry

The schematic of an asymmetric expansion is shown in Figure 1(b). As discussed earlier, asymmetry can be caused due to $\zeta \neq 0$ or unequal inclinations of walls after the expansion or a combination of the two.

Initially, simulations are carried out for parallel walls ($\alpha=0$) for three different values of perturbation (so-called geometric perturbation to the symmetric wall), i.e. $\zeta =0.02, 0.04$ and $0.06$. The streamline plots of the flow field are shown in Figure 10(a) for Re= 40, 80 and 187 and fixed value of $\zeta =0.02$. For asymmetric cases of this type, the primary recirculation bubble near the smaller step is observed to be always smaller than the other recirculation bubble. In order to substantiate this, different flow configurations with smaller step height on top and bottom walls for varying Re are studied and the results are shown in Fig 10(b). Flow profile with larger recirculation bubble near the smaller step height is also used as an initial condition. This confirms that a flow, even if it is attached to the farther wall, changes direction and attaches to the closer wall. This could be noted as by giving a small asymmetry near the expansion, the direction of attachment of flow could be biased which is a much easier way when compared to methods like strong blowing as mentioned in literature [1].

The increase in the asymmetry parameter $\zeta$ does not significantly alter the flow field than what is obtained for $\zeta =0.02$. However, the subtle changes in the flow field, due to the small asymmetry, lead to variation in the bifurcation diagram, as shown in Fig 11(a). The bifurcation diagram has a shift from the zero velocity line, and the sudden bifurcation could no longer be observed as compared to symmetry case of $\alpha=0$ (Fig . 4). Thus there is no critical Reynolds number of symmetry breaking. The flow is asymmetric even for the low Reynolds number of 30. It can also be observed that the asymmetry of flow increases with the magnitude of the perturbation. The magnitude of asymmetry is comparable for $\zeta =0.02$ and $0.04$, but for $\zeta =0.06$ the magnitude is much higher. The bifurcation diagram obtained from the present simulations are compared with the results by Fearn et al. [6] and found to be in excellent agreement, in turn confirming that Fearn's experimental set-up had an asymmetry in-between $\zeta=0.02$ and $\zeta=0.04$.

The skin friction distribution along the top and bottom walls for three Re is shown in Fig 11(b). Each of different flow patterns is presented in the figure from which analysis of the recirculation lengths can be made. It is observed from the plot that the primary recirculation bubble length at the top wall is 4.06h whereas the recirculation bubble length at the bottom wall is observed to be 3.96h showing that the flow is asymmetric at Re=40 which is also nicely predicted from the bifurcation diagram. While looking at higher Re, i.e. 80, the recirculation bubble length at the top wall is observed to be 9.8h, and at the bottom wall is 3.5h; while no sign of secondary recirculation bubble is observed at this Reynolds number. Whereas for Re=187, the primary recirculation length at the top wall is 15.1h and that at the bottom wall is 4.35h. In addition, the secondary recirculation bubble is formed at the bottom wall from 12.5h to 25.5h. These results show that the effect of the asymmetry is more profound in the bifurcation diagram even when the asymmetry is of very small magnitude.

For cases of non-parallel walls, flow is simulated for different Reynolds numbers when the top wall after the expansion is given an inclination of $\alpha=1^0$, $2^0$ and $5^0$ from the horizontal. The contour plots of the flow field are shown in Figs. 12, 13 and 14. The flow is not symmetric even at lower Re since the geometry is also asymmetric for an inclination as small as $1^0$. This can be observed from the contour plots (Figs. 12(a) & 12(b)) and from the distribution of the coefficient of skin-friction along the top and bottom walls as shown in Fig 12(c). As the Re is increased to 60, the flow asymmetry increases and the flow attaches to the bottom wall parallel to the axis. At a Re of 80, the asymmetry is further increased but no secondary recirculation bubble is observed. For Re=

90 a secondary recirculation zone is visible on the bottom wall as shown in Fig. 12(c) in which the skin friction value at a distance between 14.944h and 16.148h goes to negative. Finally, at Re=100 also a stable steady result showing secondary recirculation is obtained.

In the case of α=$2^0$ (Fig. 13), the calculations are repeated and the predictions turn out to be similar to α =$1^0$ case. For Reynolds numbers 40, 60 and 80, the flow shows asymmetry with the flow attached to the lower wall. A secondary recirculation is observed only at a Re number of 90. This is evident from the distribution of skin-friction coefficient, which shows that the value goes to negative at a distance between 14.41h and19.487h. In the case of higher α=$5^0$ (Fig 14), the asymmetry is observed to be more profound as expected. Secondary recirculation regions show up at Re=60 itself between 12.207h and 15.933h. When Re is increased to 80, the secondary recirculation region starts growing from 10.73h to 19.28h. For higher Re values for $\alpha$ =$1^0$, $2^0$ and $5^0$, the flow is observed to become unsteady, hence their discussion is not included in this work.

Furthermore, simulations of $\zeta$ =0.02 for $\alpha$ =$2^0$ are carried out for Re=40 and 80, to observe the changes in flow when both types of asymmetry are present simultaneously (Fig. 15). When Re=40 results of this case is compared with $\zeta$ =0.02 & $\alpha$ =$0^0$ case (Figs. 8-9) or $\zeta$ =0 for $\alpha$ =$2^0$ case (Fig. 13), it can be clearly observed that the asymmetry has increased substantially. The larger of the recirculation bubbles attach to the sloping wall at 5.42h from the expansion; whereas the smaller one attaches at the distance of 3.62h. Similar is the case for Re=80. The larger recirculation bubble attachment at Re=80 takes place at 13.08h from expansion on the sloping wall, while the smaller recirculation length appears to be 3.45h. Also, the secondary recirculation bubble has not been formed yet for either of the Reynolds numbers, i.e. Re=40 & 80. The results of this case of asymmetry confirm the prediction that the flow will tend to attach to the closer of the walls after the expansion. As the top wall is sloping away and the bottom wall is closer to the horizontal center-line axis, the flow attaches more easily to the bottom wall resulting in the increase in the magnitude of the asymmetry.

The knowledge on geometrical perturbation might be quite useful for conditions where flow asymmetry is required at low Re. Different step height can be employed if the magnitude of asymmetry required is small and sloping wall could be employed where larger asymmetry is required. The magnitude of the asymmetry in geometry can be adjusted by adjusting the magnitude of the perturbation in geometry applied. The superimposing of the two types of asymmetry discussed here can also be used to control the flow. For example, the diverging wall on one side might give larger values of asymmetry and its fine adjusting can be done by changing the step heights. An attempt towards correlating the magnitude of geometrical perturbations and the magnitude of flow asymmetry has plenty of scope for future works.

It is also noteworthy to mention that the typical problem of stability or large grid size reported in [14] for high Re has not been encountered in the present simulations. We have successfully used uniform mesh and reasonable grid size for Re up to 187 and obtained quite accurate results with reasonable computational cost. Hence, it is not necessarily one has to go for the non-uniform grid for higher Re calculations, while the uniform mesh is quite effective in parallelizing the code, especially in the context of load balancing.

## 5. Conclusions

While flows in the symmetric sudden expansion are well investigated, present research involves both conventional CFD and lattice Boltzmann simulation of flow through slightly asymmetric and diverging sudden expansion geometries. The major conclusions from the studies are as follows:

1. Comparison with available literature shows that the Lattice Boltzmann technique accurately captures bifurcating flow structures and critical Re in symmetric sudden expansion while compared to the N-S simulations.

2. Comparison between the symmetrically-diverging and parallel walls shows that increasing wall-divergence reduces the critical Re.

3. The variation in step heights brings asymmetry in the system while maintaining the parallel wall structure. Such geometry still develops bifurcating flow without a sharp critical Re. The variation in recirculation patterns, however, becomes more prominent with increasing Re.

4. When one wall diverges, keeping another wall parallel to the incoming flow direction, the flow ceases to bifurcate and the short recirculation attaches to the parallel wall. In this case, variation in step height indicates short recirculation always forms on the wall closer to the inlet. This conclusion would be particularly helpful in biasing the direction of attachment of flow or adjusting the reattachment lengths in a much easier way.

**Acknowledgements**

Simulations are carried out on the computers provided by the Indian Institute of Technology Kanpur (IITK) (www.iitk.ac.in/cc) and the manuscript preparation as well as data analysis has been carried out using the resources available at IITK. This support is gratefully acknowledged. Discussion with Dr. A. K. Saha (IIT Kanpur) is also gratefully acknowledged.

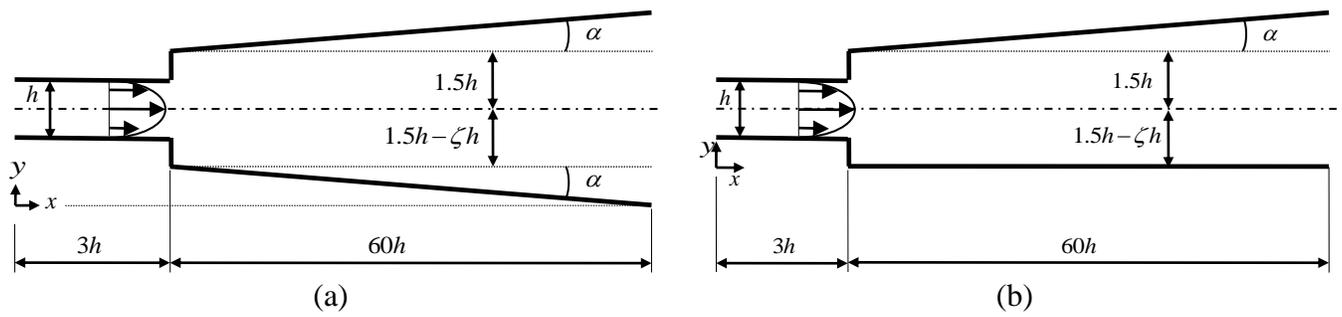

Fig 1. (a) Symmetric sudden expansion in a diverging channel in which each wall is inclined at an angle α with the inflow direction. (b) Asymmetric sudden expansion: Asymmetry can be due to different step sizes of the expansion or due to non-parallel walls or both.

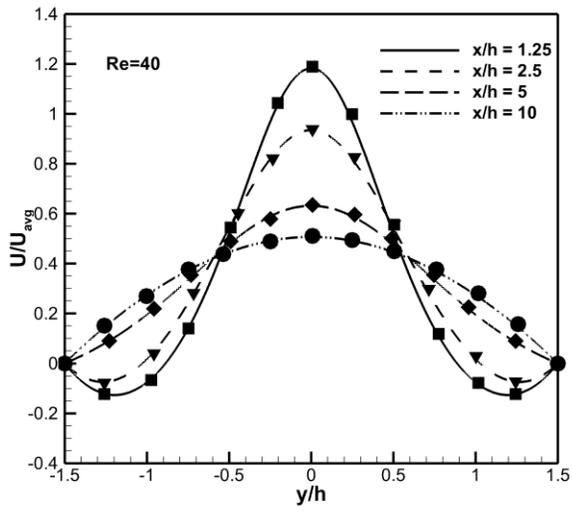
(a)
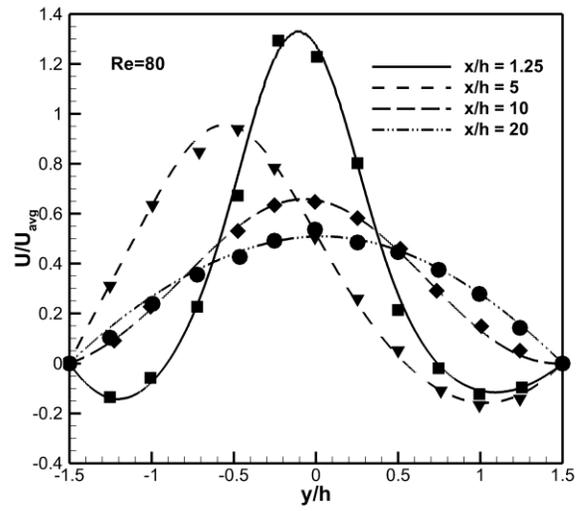
(b)
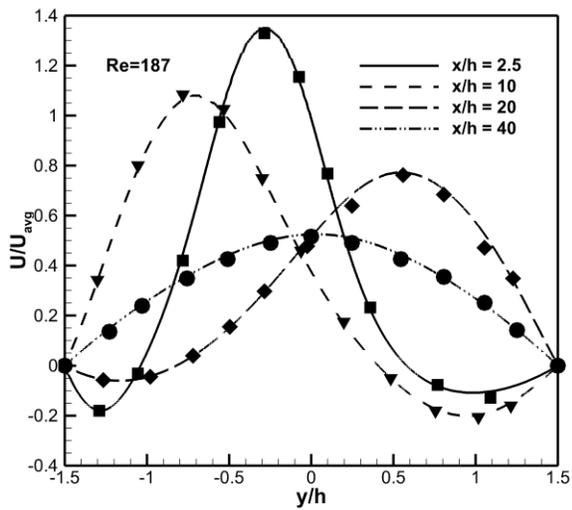
(c)
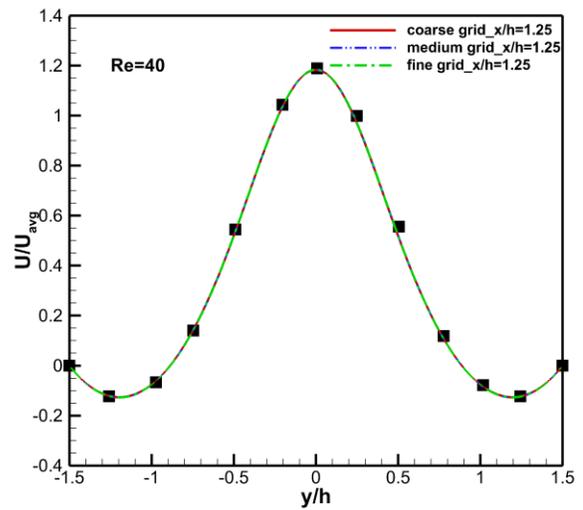
(d)

Fig 2. Velocity profile validation for symmetric sudden expansion using N-S simulations for α=0, at (a) Re=40 (with Hawa and Rusak [8]), (b) Re=80 and (c) Re=187 (with Battaglia et al [29]). Here, U represents the velocity in the x direction. Also, in this figure, y varies from -1.5 to 1.5 to comply with the representation given in the references with which the results are compared. (d) A grid independence test done on symmetric geometry with α=0 and Re=40

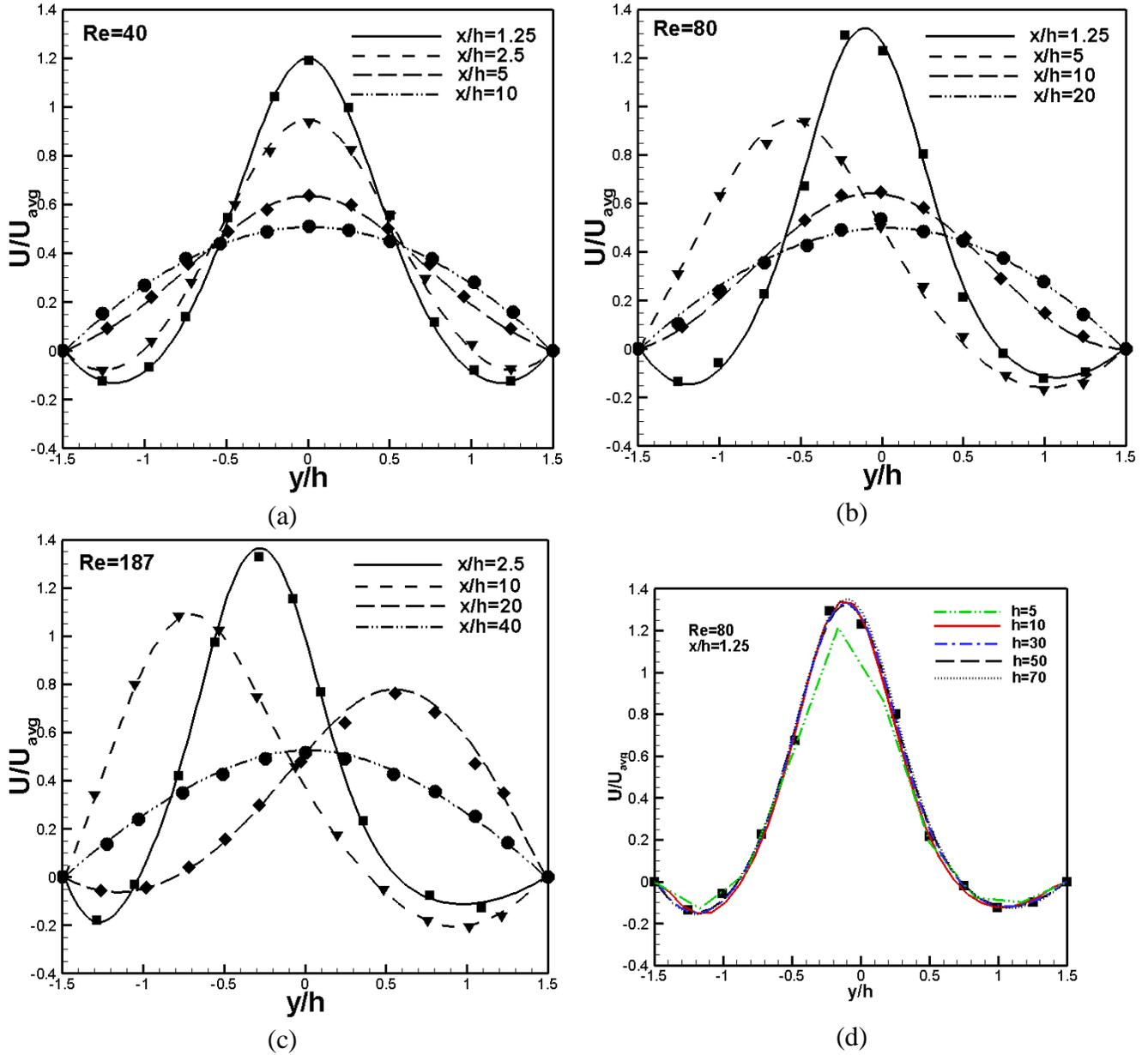

Fig 3. Velocity profile validation for symmetric sudden expansion using LBM for α=0, at (a) Re=40 (with Hawa and Rusak [8]), (b) Re=80 and (c) Re=187 (with Battaglia et al [29]). Here, U represents the velocity in the x direction. Also, in this figure, y varies from -1.5 to 1.5 to comply with the representation given in the references with which the results are compared. (d) A grid independence test done on symmetric geometry with α=0 and Re=80

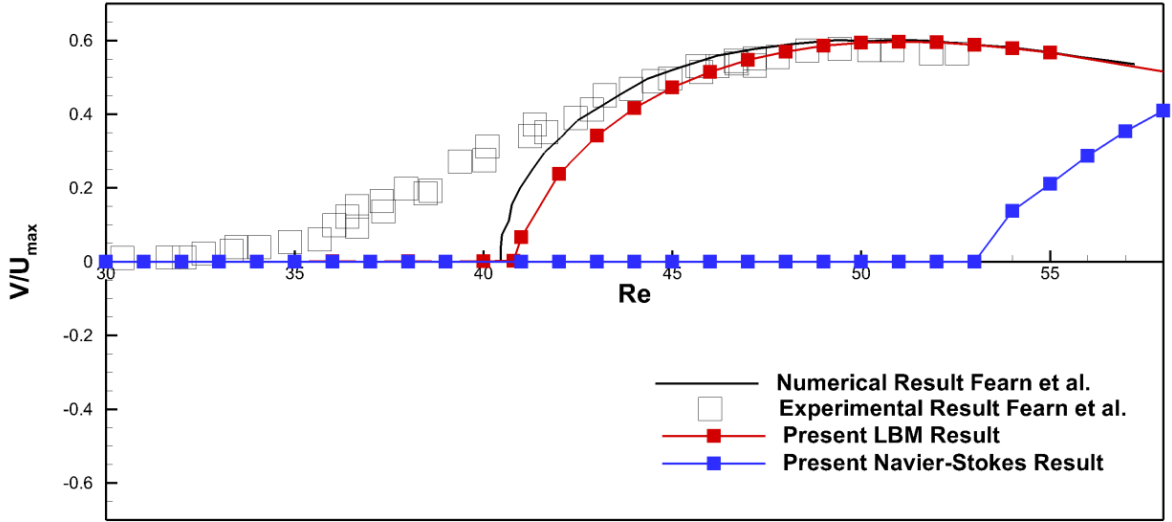

Fig 4. Bifurcation diagram for α=0 compared with results from Fearn et al. [6]. V is the velocity component in the y direction at the point on the centerline of the channel at a distance of 6.38h from the plane of expansion. $U_{max}$ is the maximum velocity at the inlet.

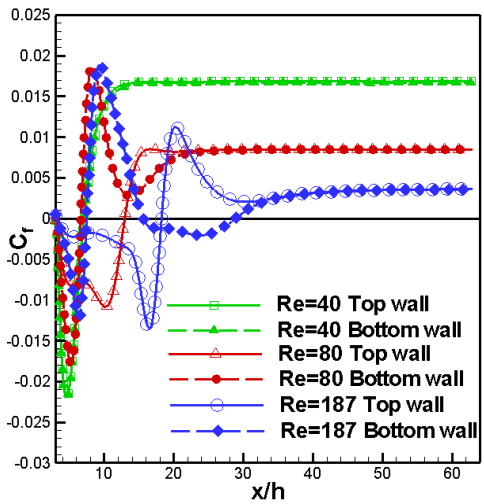

(a)

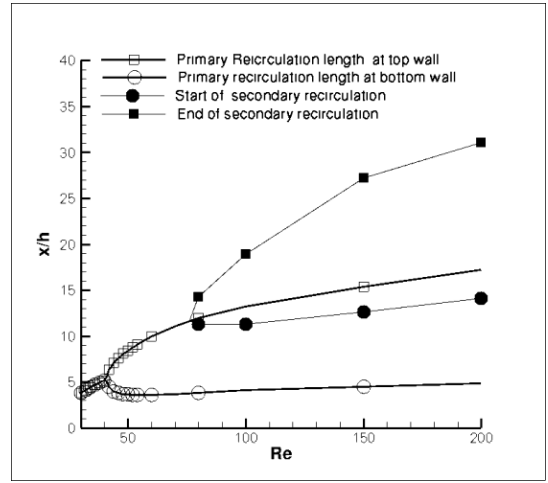

(b)

Fig 5. (a) Distribution of skin friction coefficient along the top and bottom walls for the case of α=0. (b) Plot showing the variation of recirculation lengths with Re for α=0. The recirculation lengths are measured from the plane of expansion.

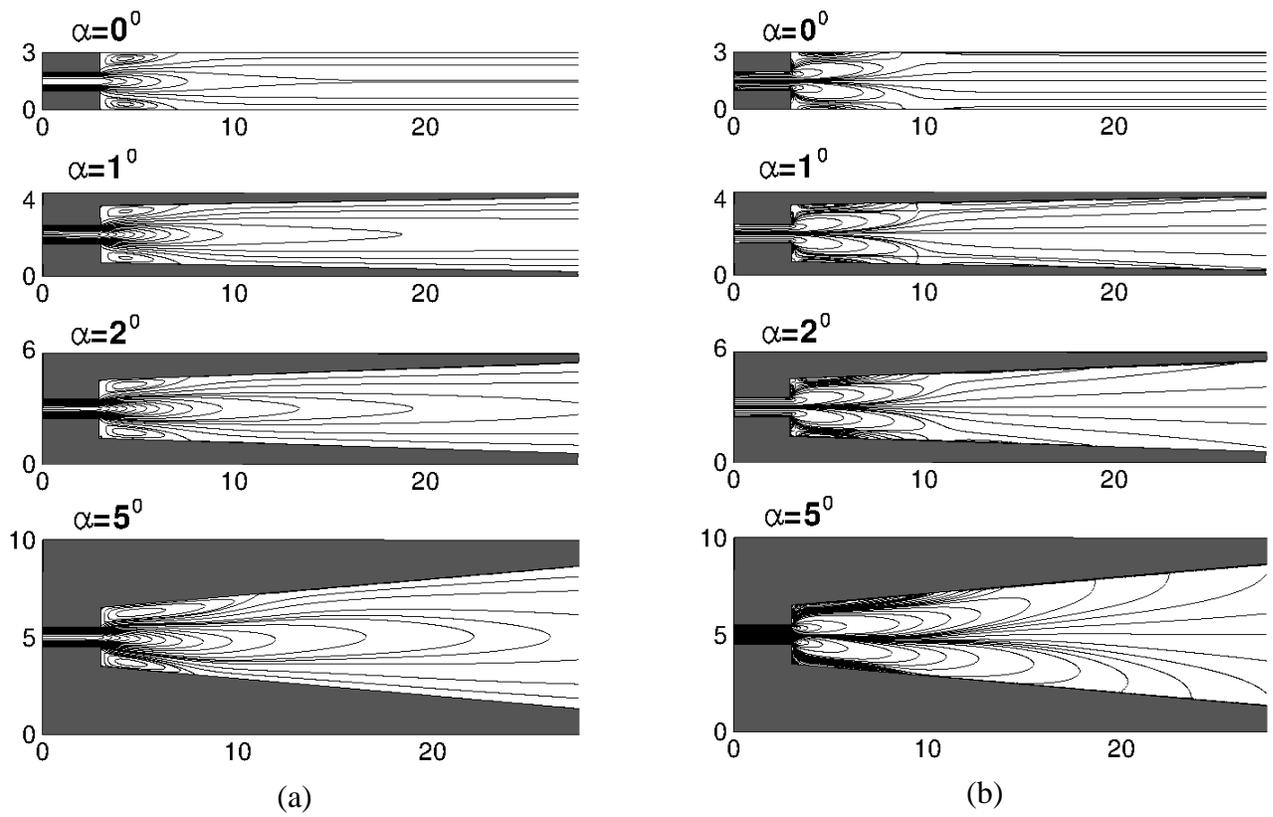

Fig 6. Streamlines (a) and vorticity contour (b) for flow in symmetric sudden expansion at Re =40. After the expansion, each wall diverges at an angle $\alpha$ with respect to the inflow direction

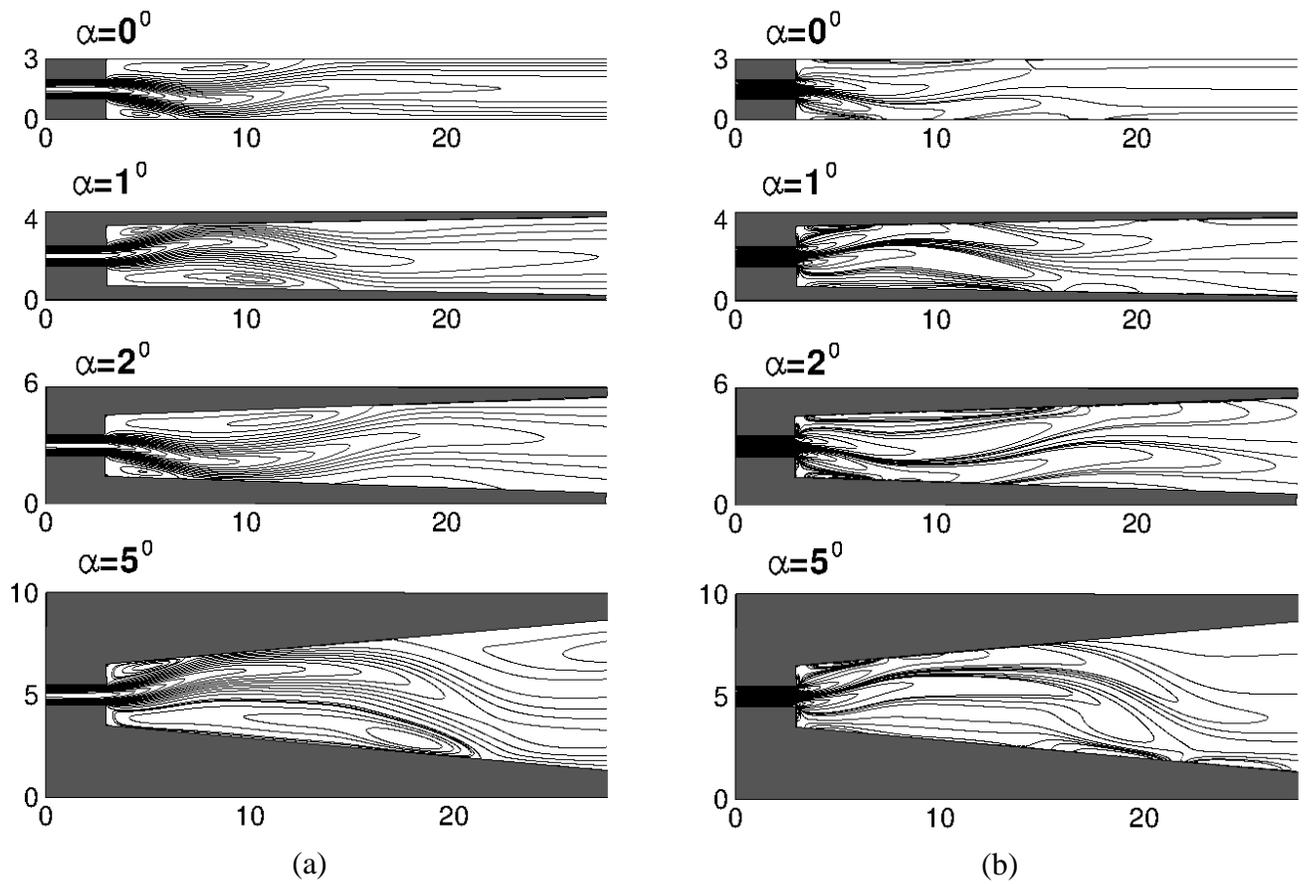

Fig 7. Streamlines (a) and vorticity contour (b) for flow in symmetric sudden expansion at Re =80. After the expansion, each wall diverges at an angle $\alpha$ with respect to the inflow direction

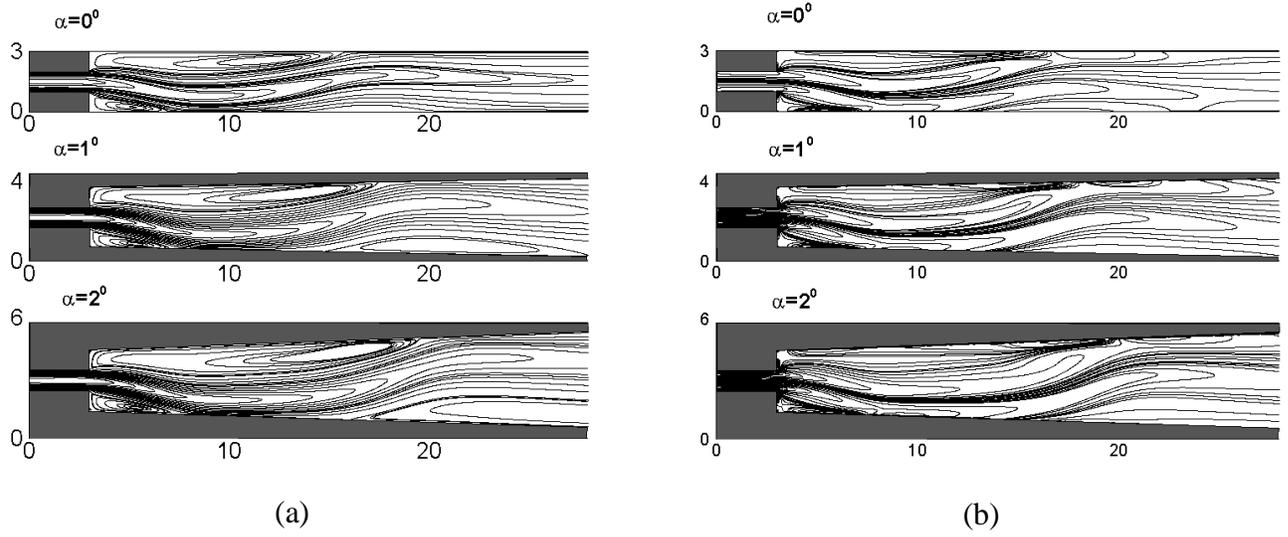

Fig 8. Streamlines (a) and vorticity contour (b) for flow in symmetric sudden expansion at Re =120. After the expansion, each wall diverges at an angle $\alpha$ with respect to the inflow direction

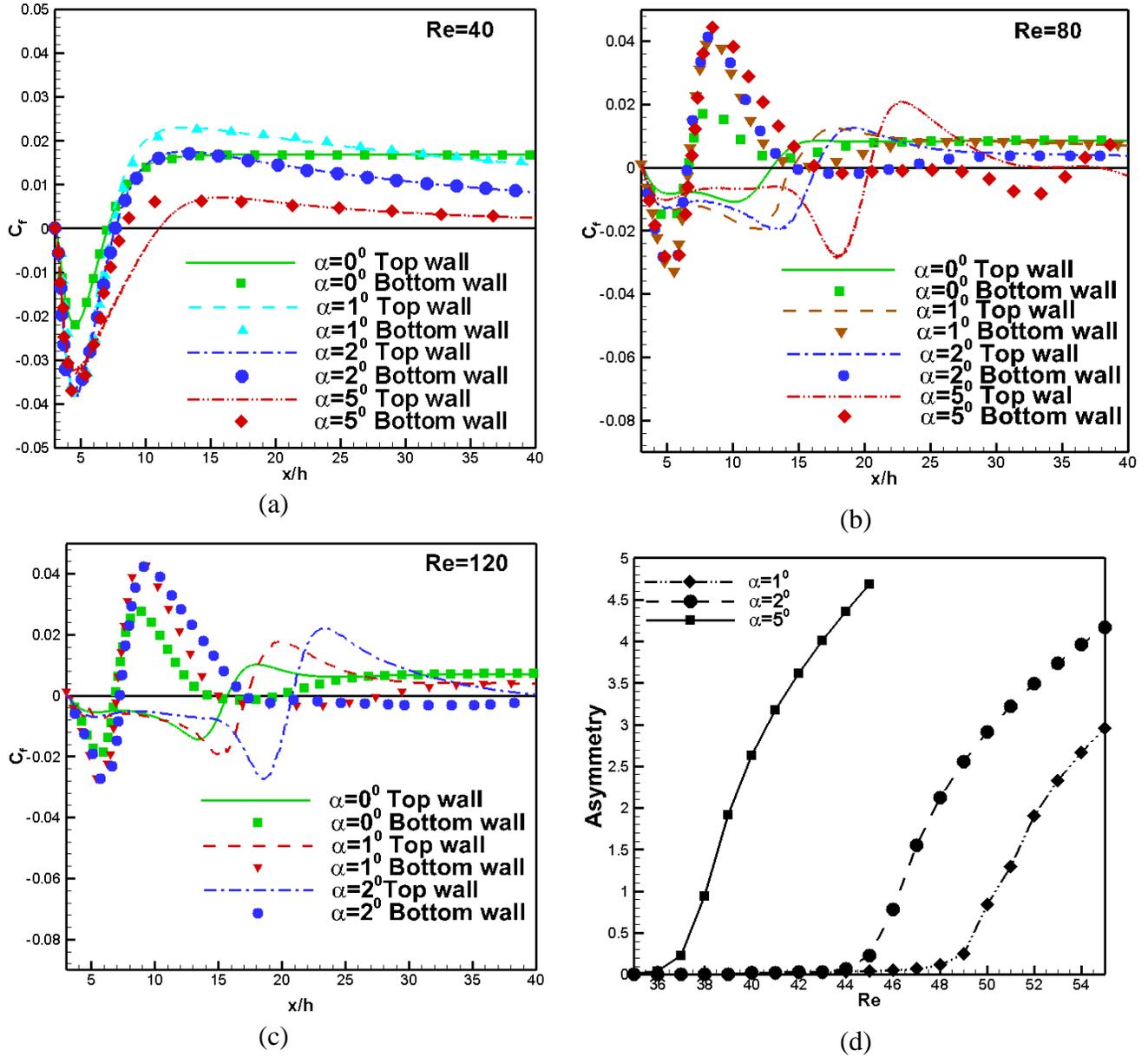

Fig 9: Distribution of skin friction coefficient along the top and bottom walls for (a) Re=40, (b)Re=80 and (c)Re=120. Solid lines and symbols show $C_f$ distribution along top wall and bottom wall respectively. (d) Symmetry breaking of flow with Re for different α. Here Re is defined in terms of average inlet velocity and inlet section height. Y axis is the sum of absolute vertical velocity at each point on the centerline of the channel.

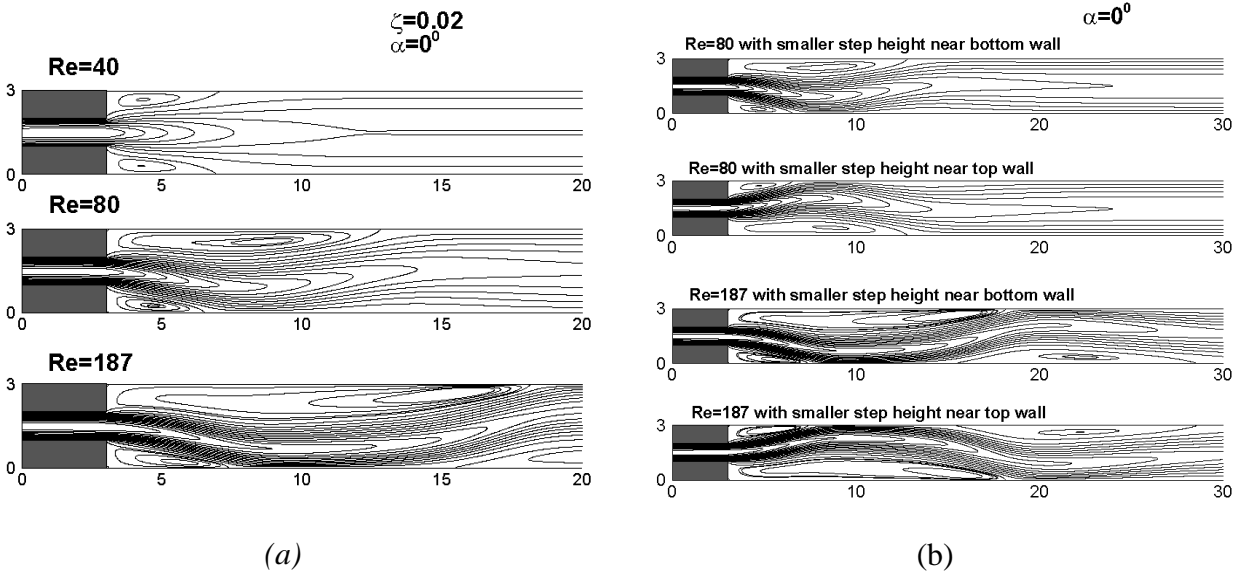

Fig 10: (a) Velocity contour plots for α =00 and $\varsigma$ =0.02 (b) Velocity contour plots to show that the flow will always attach to the closer of the two bounding walls.

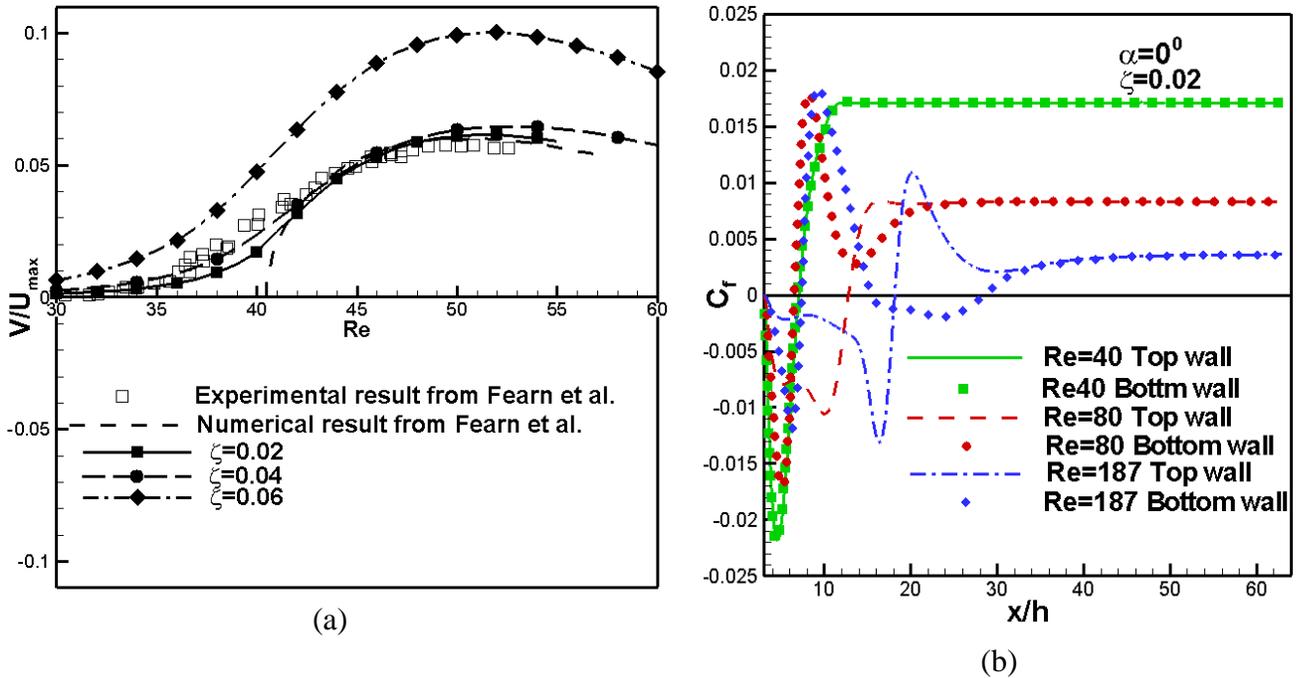

Fig 11: Bifurcation diagram for different $\varsigma$ values at α =$0^0$. The results from Fearn et al. is for $\varsigma$ =0. (b) Distribution of skin friction for α=$0^0$ and $\varsigma$ =0.02 for different Re. Solid lines and the dotted lines show $C_f$ distribution along top wall and bottom wall respectively.

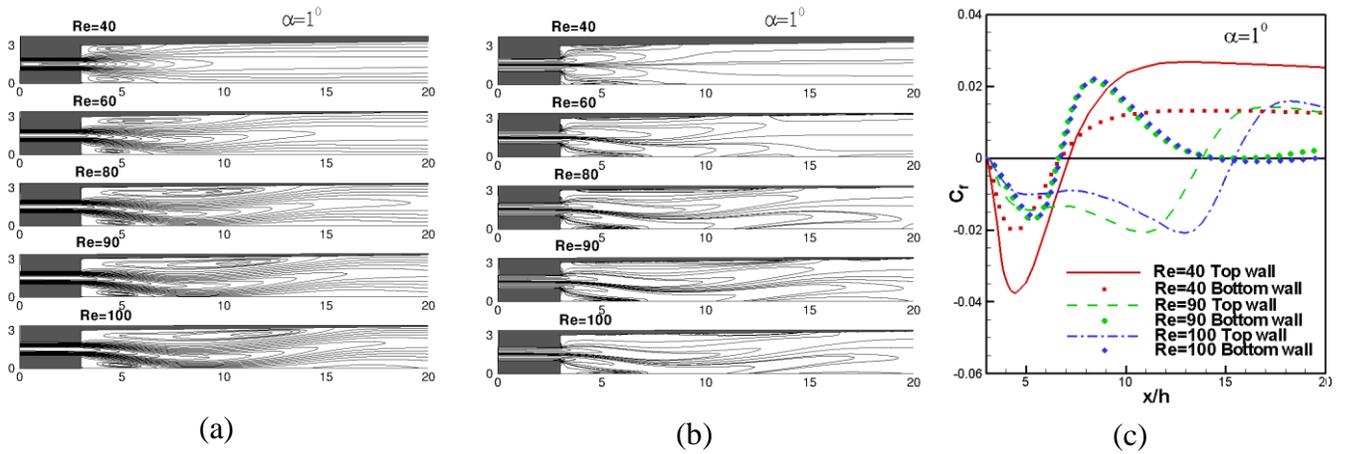

Fig 12: Results of α =1⁰ for $\varsigma$ =0 at different Re. (a) velocity contours (b) vorticity contours and (c) Skin friction distribution. Solid lines and the dotted lines show $C_f$ distribution along top wall and bottom wall respectively.

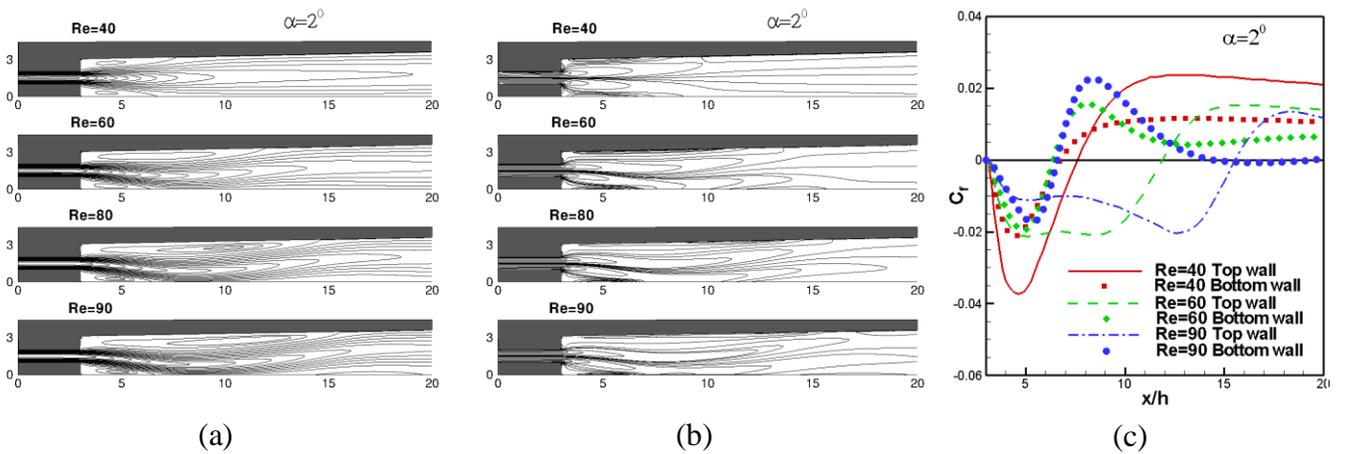

Fig 13: Results of α =2⁰ for $\varsigma$ =0 at different Re. (a) velocity contours (b) vorticity contours and (c) Skin friction distribution. Solid lines and the dotted lines show $C_f$ distribution along top wall and bottom wall respectively.

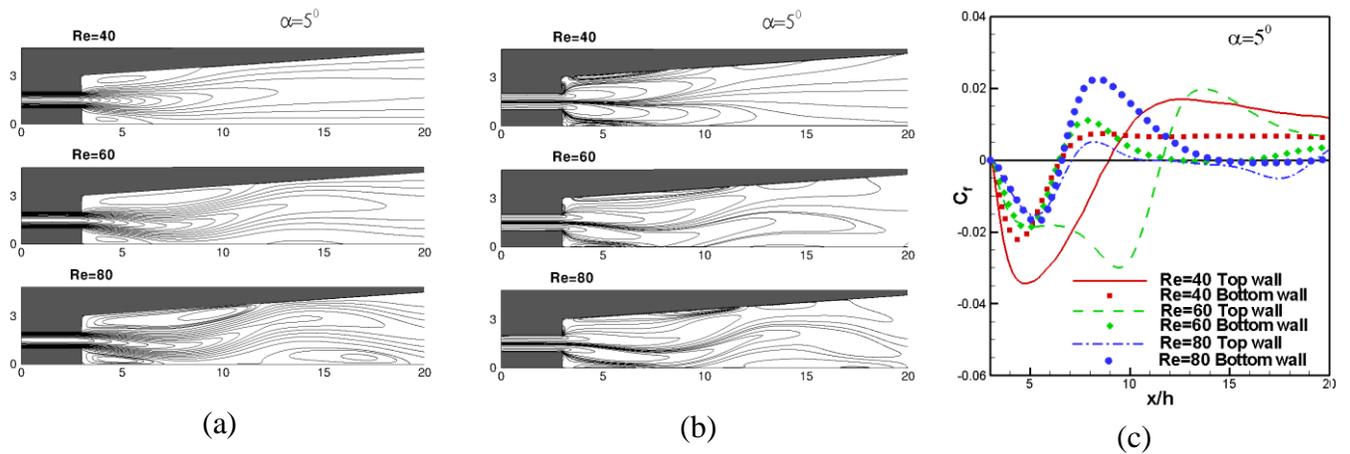

Fig 14: Results of α =5$^0$ for $\varsigma$ =0 at different Re. (a) velocity contours (b) vorticity contours and (c) Skin friction distribution Solid lines and the dotted lines show C$_f$ distribution along top wall and bottom wall respectively.

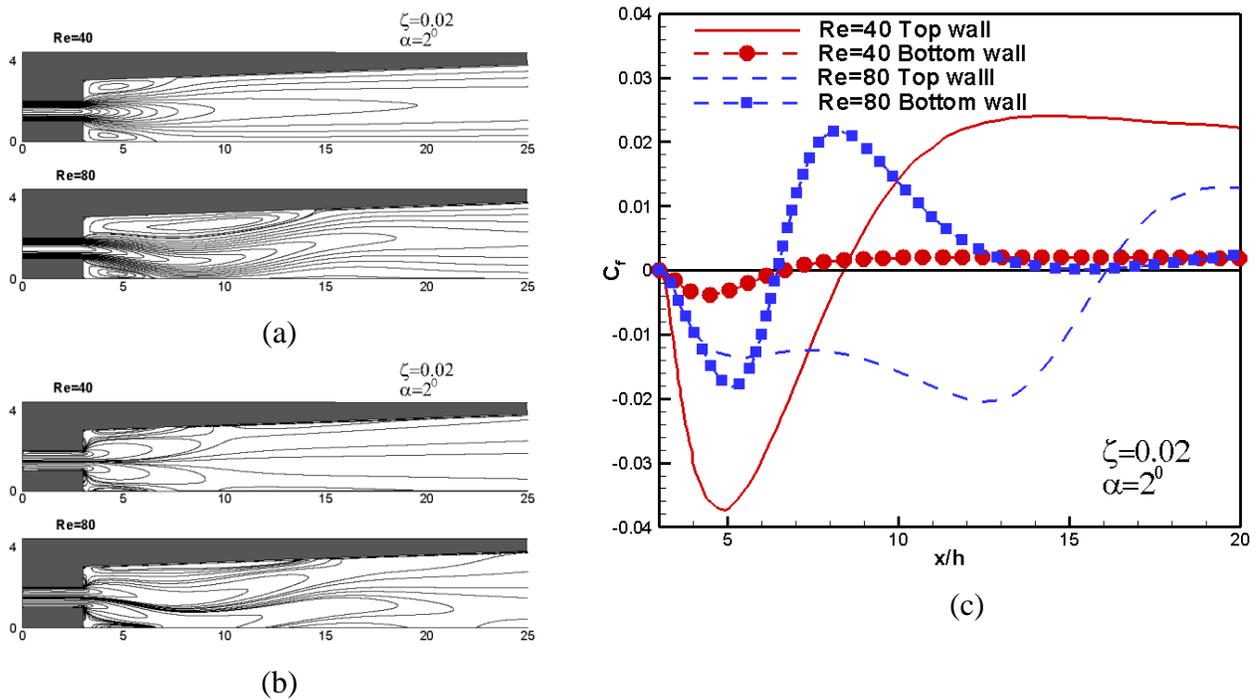

Fig 15: Results for geometry with the superimposition of two types of asymmetry. α =2$^0$ for $\varsigma$ =0.02. (a) velocity contours (b) vorticity contours and (c)Skin friction distribution.